\documentclass{aa}
\usepackage{times}
\usepackage{array}
\usepackage{epsfig}
\usepackage{amssymb}
\usepackage{amsmath}
 \def\mso{\,\mathrm{M}_\odot}                                                  
 \def\rso{\,{\rm R}_\odot}

 \def\teff{\log\, T_{\rm eff}\,}                                                
 \def\llso{\log\, L/{\rm L}_\odot \,}                                         

 \def\h1{\hangindent=1.0truecm \hangafter=0}                                    
 \def\simle{\mathrel{\hbox{\rlap{\hbox{\lower4pt\hbox{$\sim$}}}\hbox{$<$}}}}    
 \def\simgr{\mathrel{\hbox{\rlap{\hbox{\lower4pt\hbox{$\sim$}}}\hbox{$>$}}}}    
 \def\msoy{\, \mso~{\rm yr}^{-1}}

 \def\hb{\hfill\break}
 \def\h2{\hb\noindent \hangindent=0.5cm \hangafter=1}

 \def\utw{\smash{\rlap{\lower5pt\hbox{$\sim$}}}}
 \def\udtw{\smash{\rlap{\lower6pt\hbox{$\approx$}}}}

\newcommand{\Msun}{\mso}
\newcommand{\Rsun}{\rso}

\begin{document}
\thesaurus{ 06 (08.03.4; 08.05.3; 08.18.1; 08.19.5)}

\title{Implications of massive close binaries for black hole
  formation and supernovae}

\author{S. Wellstein \and N. Langer}

\institute{
  Institut f\"ur Physik, Universit\"at
  Potsdam, Am Neuen Palais 10, D--14415~Potsdam, Germany
}

\offprints {S. Wellstein (email: {\tt stephan@astro .physik.uni-potsdam.de});
  or N. Langer (email: {\tt ntl@astro .physik.uni-potsdam.de})}

\date{Received  ; accepted , }

\maketitle

\begin{abstract}
  The progenitor evolution of the massive X-ray binary Wray~977 is
  investigated using new models of massive close binary evolution.
  These models yield constraints on the mass limit for neutron
  star/black hole formation in single stars, $M_{\rm BH}$. We argue
  for quasi-conservative evolution in this system, and we find $M_{\rm
    BH} > 13..21\Msun$ from the existence of a neutron star in
  Wray~977, with the uncertainty being due to uncertainties in the
  treatment of convection. Our results revise earlier published much
  larger values of $M_{\rm BH}$ derived from the parameters of
  Wray~977.
  
  Then, on the basis of a grid of 37 evolutionary models for massive
  close binaries with various initial masses, mass ratios and periods,
  we derive primary initial-final mass, initial mass-final helium core
  mass, and initial mass-final CO-core mass relations for the various
  mass transfer cases of close binary evolution.  From these models we
  derive for single stars that $M_{\rm BH} \simle 25\mso$, independent
  of whether most black hole binaries formed through the Case~A/B or
  the Case~C binary channel.  Using our grid of binary models, we
  obtain a consistent scenario for the formation of black holes in
  binary systems.
  
  We emphasize that in binaries the critical initial mass limits for
  neutron star/black hole formation and for white dwarf/neutron star
  formation are very different from the corresponding values in single
  stars. While the first may well be above 100$\mso$ in Case~A/B
  binaries, the latter is found to be in the range 12...15$\mso$
  instead of the canonical value of 8...10$\mso$ usually quoted for
  single stars. This effect should not be neglected in population
  synthesis studies of massive binary systems.  Also, neutron star and
  black hole mass functions obtained for single stars can not per se
  compared to the masses of compact objects in binary systems.
  
  Massive close binaries produce also Type~Ib and~Ic supernovae.  We
  find two different types of supernova progenitor structure in our
  models, one with remaining helium masses of the order of 1$\mso$
  which stems from an intermediate progenitor initial mass range
  (about 16...25$\mso$), and another with one order of magnitude
  smaller remaining helium masses from initial masses above and below
  this. A possible connection to the distinction of Type Ib and
  Type~Ic supernovae, and implications from the Type Ic supernova
  SN1998bw and its associated $\gamma$-ray burst are discussed.

\end{abstract}

\section{Introduction}

The evolution of stars is considerably complicated by the presence of
a close companion, which may lead to either extreme mass loss, to
strong mass accretion, or even to the merging of both stars. An
overview of the evolutionary possibilities in massive close binaries
is given by Podsiadlowski et al. (1992), while evolutionary models for
parts of the --- unfavorably large --- initial parameters space of
binaries have been computed, e.g., by Paczinsky (1967), Kippenhahn
(1969), de Loore \& De Greve (1992), Pols (1994); see also Vanbeveren
(1998ab, and references therein).  On the other hand, close binaries
provide unique possibilities to test and constrain uncertainties
inherent in the evolution of stars in general.

An example is given by Ergma \&\ van den Heuvel (1998), who showed
that massive close binary systems containing a compact stellar remnant
(neutron star or black hole) can constrain the initial mass limit for
black hole formation $M_{\rm BH}$.  In the present paper, we want to
pursue this idea in a quantitative way. In principle, the problem is
simple: a valid progenitor model for a system containing a neutron
star yields the initial mass of the neutron star progenitor and thus a
lower limit to $M_{\rm BH}$, and systems containing a black hole yield
an upper limit to $M_{\rm BH}$.  This procedure is not hopeless, even
though the calculations of the progenitor evolution of the observed
systems can not provide the information whether a model component in
its final stage evolves into a neutron star or a black hole: The
observed orbital period together with the properties of the normal
star in the system may allow only for a very limited range of initial
masses for the progenitor of the compact component.  On the other
hand, all the uncertainties involved with the theoretical description
of mass transfer in massive close binaries enter the problem and
increase the error bar on the progenitor masses.

Note that $M_{\rm BH}$ is not a well defined quantity in binary
systems; i.e., whether a star in a binary system forms a black hole or
a neutron star depends not only on its initial mass but also on its
evolutionary history (see below). We assume implicitly in this paper
that $M_{\rm BH}$ is in fact well defined for single stars.  This is
also not guaranteed. E.g., the initial rotation rate may be a
parameter to be considered in addition to the initial mass (cf., Heger
et al. 1999). Ergma \& van den Heuvel (1998) even argue from the
properties of massive close binaries with compact companions that in
single stars, $M_{\rm BH}$ must be depending on a second parameter.
However, even though we can not show the contrary, we attempt here to
disprove their main argument for this.

Two criteria may help to chose those observed systems which are best
suited to constrain $M_{\rm BH}$. First, if one would pick systems for
which one could assume that the system progenitor evolution proceeded
more or less conservatively, i.e. that most of the system initial mass
remained in the system, one could avoid the huge uncertainties
inherent in theories of mass outflow from the system, either through
the second Lagrangian point or in the course of a common envelope
evolution (Podsiadlowski et al. 1992). Since strong mass outflow
efficiently removes angular momentum and thus results in short
periods, one should avoid the systems with the shortest periods.
Second, it would be most efficient to investigate systems of which one
can hope that the compact star's progenitor mass is close to $M_{\rm
  BH}$. For systems containing a neutron star, this means one should
look for the most massive systems.

The massive X-ray binary Wray 977/GX~301-2 (4U 1223-62) fulfills both
criteria.  It contains an X-ray pulsar (GX~301-2) and the B supergiant
Wray 977 (BP Cru).  The latest and most reliable determination of the
stellar parameters of Wray~977 has been performed by Kaper \&\ Najarro
(1999).  They found log ${\mathrm T_{eff}/[K] = 4.23...4.30}$ and a
radius of $R = 60...70 \Rsun$, with preferred values of 4.23 and 62
$\Rsun$. A radius of 62 $\Rsun$ combined with the empirical mass
function of Wray~977/GX 301-2 (Kaper et al. 1995) and the absence of
eclipses implies a lower mass limit for the B star of $M=39...40
\Msun$.  An independent lower mass limit for Wray~977 of $\sim 40\mso$
has been derived spectroscopically from the velocity amplitude by
Kaper \&\ Najarro (1999). Constraints on the properties of Wray 977
obtained by Koh et al. (1997) agree with these values.  The observed
period of the binary is 44.15 days.

Ergma \&\ van den Heuvel (1998) concluded that the neutron star
progenitor mass was initially larger than the present mass of the
B~star.  Here, we present massive close binary models from zero age
until beyond the death of the primary component (defined here as the
initially more massive star in the system), which constrain the
progenitor evolution of Wray 977/GX~301-2. After describing our
computational method in Section~2, we present our best model for Wray
977/GX~301-2's progenitor evolution in Section~3, which results in an
initial mass for GX~301-2 of only 26$\mso$.  In Section~4, we discuss
the evolution of the stellar, the helium core, and the CO-core masses
of primaries in massive binary system on the basis of a new grid of
computed systems.  In Section~5, our grid of binary models is used to
derive the transformation of the initial mass limit for neutron star
formation obtained from binaries to the single star case. In Section~6
we analyze the evolution of the chemical structure of massive
primaries and implications for Type~Ib/c supernovae. In Section~7 we
summarize our results to a global picture of massive close binary
evolution and a scenario for the formation of black hole and neutron
star binaries.

\section{Computational methods}

We computed the evolution of massive close binary systems using a
computer code generated by Braun (1997) on the basis of an implicit
hydrodynamic stellar evolution code for single stars (cf. Langer 1991,
1998). It invokes the simultaneous evolution of the two stellar
components of a binary and computes mass transfer within the Roche
approximation (Kopal 1978).  The entropy of the accreted material is
assumed to be equal to that of the surface of the secondary star,
where gravitational energy release due to mass accretion is treated as
in Neo et al. (1977); see also Braun \& Langer (1995).  Even though
mass loss due to stellar winds is included for both components (see
below) --- with a corresponding angular momentum loss according to
Brookshaw \& Tavani (1993) --- the present calculations deal only with
contact-free evolutionary stages, and therefore no other source of
mass outflow from the system is included.  Our models can thus be
called quasi-conservative.

We use standard stellar wind mass loss rates, i.e., the rate of
Nieuwenhuijzen \& de Jager (1990), except for hot stars.  For OB stars
($\teff > 15\, 000\,$K), we use the theoretical radiation driven wind
models of Kudritzki et al. (1989) with wind parameters $k=0.085$,
$\alpha = 0.657$, $\delta = 0.095$, and $\beta =1$ (Pauldrach et al.
1994).  I.e., the dependence of the mass loss rate on the luminosity,
effective temperature, mass, and surface hydrogen mass fraction is
taken into account. For hydrogen-poor stars, i.e. stars with a surface
hydrogen mass fraction $X_{\rm s} < 0.4$, we used a relation based on
the empirical mass loss rates of Wolf-Rayet stars derived by Hamann et
al. (1995) for massive stars $\llso \ge 4.5$ and by Hamann et al.
(1982) for helium stars in the range $3.5 \le \llso \le 4.5$, i.e.
\[
\log(-\dot M_{\mathrm WR}/(\msoy))=
\]
\begin{equation} \hfill
 \left\{ \begin{array}{ll}
  \displaystyle -11.95+1.5\llso-2.85 X_{\rm s}&\hbox{\ for\ }
  \llso\ge 4.5\\ \rule{0pt}{5mm}-35.8+6.8\llso&\hbox{\
  for\ } \llso<4.5\\ \end{array} \right.
\end{equation}

Note that for core helium burning helium stars, this equation gives
mass loss rates very close to that proposed by Langer (1989).  To
account for recent revisions of empirical Wolf-Rayet mass loss rates
by Hamann \& Koesterke (1998), who suggested that previously derived
values for massive Wolf-Rayet stars may overestimate the mass loss by
a factor of 2...3, we also computed sequences where this rate has been
multiplied by a factor 0.5, i.e.,
\[
\log(-\dot M_{\mathrm WR}/(\msoy))=
\]
\begin{equation} \hfill
 \left\{ \begin{array}{ll}
  \displaystyle -12.25+1.5\llso-2.85 X_{\rm s}&\hbox{\ for\ }
  \llso\ge 4.45\\ \rule{0pt}{5mm}-35.8+6.8\llso&\hbox{\
  for\ } \llso<4.45\\ \end{array} \right.
\end{equation}

Convection and semiconvection have been treated according to Langer et
al. (1983), mostly using a semiconvective efficiency parameter of
$\alpha_{\mathrm{sc}} = 0.01$ (cf. also Braun \& Langer 1995).
Time-dependent thermohaline mixing is followed in a diffusion scheme
according to Wellstein et al. (1999; cf. also Braun 1997), which is
based on an analysis of Kippenhahn et al. (1980).

Opacities are taken from Iglesias \& Rogers (1996).  Changes in the
chemical composition are computed using a nuclear network including
the pp-chains, the CNO-tri-cycle, and the major helium-, carbon and
oxygen burning reactions. For the $^{12}$C($\alpha$,$\gamma$)$^{16}$O
nuclear reaction rate we followed Weaver \& Woosley (1993) and used a
value of 1.7~times that of Caughlan \& Fowler (1988).  Further details
about the computer program and input physics can be found in Langer
(1998) and references therein. All models in this work use an
approximately solar initial chemical composition with a mass fraction
of all elements heavier than helium (``metals'') of $Z=0.02$.  The
mass fractions of hydrogen and helium are set to $X=0.7$ and
$Y=1-X-Z=0.28$, respectively.  The abundance ratios of the isotopes
for a given element are chosen to have the solar system meteoritic
abundance ratios according to Grevesse \& Noels (1993).

\section{The smallest possible initial mass of GX 301-2}

The derivation of initial masses of neutron star progenitors in
binaries can only yield lower limits to $M_\mathrm{BH}$. These limits
are immediately valid for single stars, since the stronger mass loss
in primaries of close binaries compared to that in single stars can
cause more massive primaries than single stars to form neutron stars,
but not vice versa. As due to the uncertainties in the binary
evolution models one always ends up with a possible range of initial
masses for the neutron star progenitor in a given system, only the
smallest mass in this range leads to a stringent constraint on
$M_\mathrm{BH}$.  Therefore, we attempt to find the lowest possible
initial mass for the progenitor of the neutron star GX~301-2.

\subsection{Restricting the initial system parameters}

\begin{figure*}[ht]
\begin{centering}
\epsfxsize=0.8\hsize
\epsffile{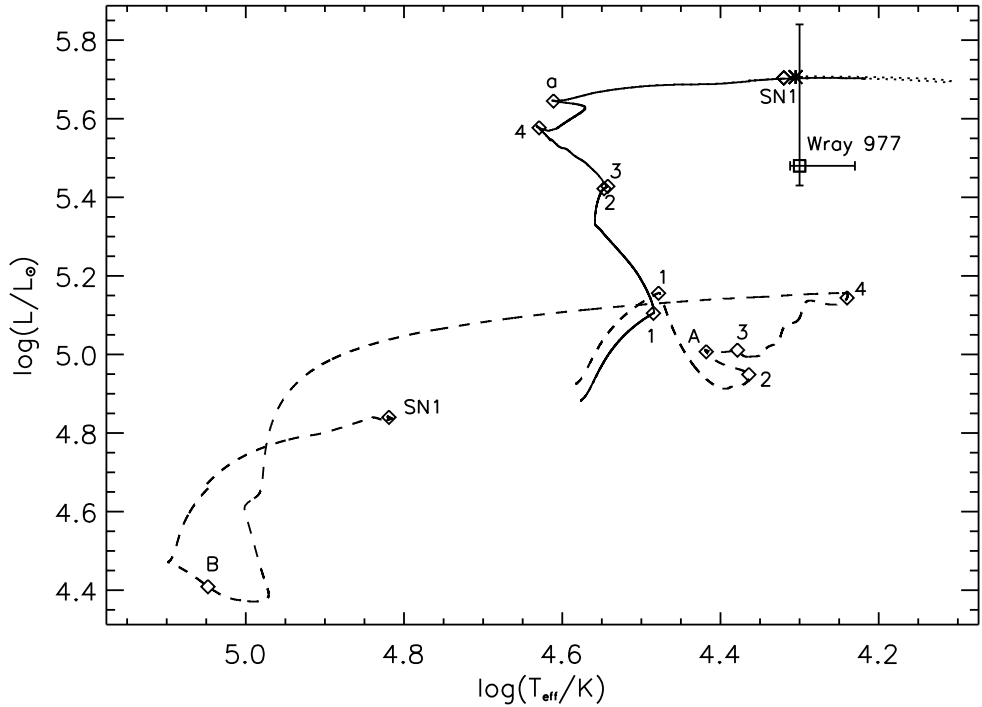}
\caption{Evolution of the components of our progenitor
  model for Wray~977/GX~301-2 (model No.~8; cf.
  Table~\ref{fig:systems}) in the HR diagram.  The dashed line
  corresponds to the evolution of the primary from the zero age main
  sequence to its supernova explosion.  The solid line corresponds to
  the evolution of the secondary from the zero age main sequence until
  the onset of reverse mass transfer onto the neutron star (solid
  line); the dotted continuation of the solid line designates the
  secondary's evolution up to its supernova explosion (marked by a
  asterisk) if the neutron star companion were absent.  The end of the
  solid line corresponds to a central helium mass fraction of the
  secondary of $Y_{\rm c} = 0.14$.  Beginning and end of the various
  mass transfer phases are marked as follows. 1: begin of Case~A, 2:
  end of Case~A, 3: begin of Case~AB, 4: end of Case~AB.  A/a
  designates core hydrogen exhaustion for the primary/secondary, B the
  end of primary's core helium burning. The position of both
  components at the time of the primary's supernova explosion is
  marked by a diamond and labeled `SN1'. The position of Wray~977
  according to Kaper \& Najarro (1999) is indicated (square) together
  with the error bars.  }
\label{fig:hrd}
\end{centering}
\end{figure*}

Since the neutron star progenitor evolved first into a supernova, the
B~star Wray~977 is very likely the initially less massive star in the
binary system --- we thus designate it here as the secondary
component. Note that even though a reversal of the supernova order ---
i.e., the secondary component exploding earlier than the primary ---
is possible (e.g., Pols 1994, or System No.~10 in Section~4 below),
this assumption would lead to larger neutron star progenitor masses
than the assumption of the normal supernova order.  If we define, as
usual, $\beta$ as the fraction of the mass transfered from the primary
to the secondary which remains in the binary system, i.e. the fraction
$1-\beta$ is ejected from the system, one can easily see that
$\beta\rightarrow 1$ (i.e. conservative evolution) leads to the
smallest possible initial masses for the neutron star progenitor.
E.g., assuming $\beta =0$ implies that the progenitor mass of GX~301-2
must be larger than Wray~977's present mass, i.e., $M_{\rm 1,i} \simgr
40\mso$ (cf. Ergma \& van den Heuvel 1998), while every solar mass
which is successfully transfered form the neutron star progenitor to
Wray~977 allows the initial mass of GX~301-2 to have been roughly one
solar mass smaller.

Furthermore, smaller neutron star progenitor masses can be achieved
the closer the initial mass ratio $q$ is to one. Keeping the total
mass $M$ in the system constant up to the first supernova explosion
($\beta =1$), the smallest possible initial mass of the neutron star
progenitor is $M/2$, which corresponds to $q=1$.

Finally, we need to know the initial period of our most constraining
binary model. There are two reasons which made us prefer a Case~A
evolution (mass transfer during core hydrogen burning). First, simple
estimates --- which are possible and meaningful for conservative
systems --- show that initial periods corresponding to Case~A result
in final periods which are in the range of that found in
Wray~977/GX~301-2 (Pols 1994).  Second, Case~A mass transfer yields
smaller values for $M_{\rm BH}$ in single stars than Case~B or~C
\footnote{as Kippenhahn \& Weigert (1967) and Podsiadlowski (1992), we
  define the Cases~A, B, and C evolution corresponding to mass
  transfer during core hydrogen burning, after core hydrogen burning
  but before core helium exhaustion, and after core helium exhaustion,
  respectively} (as will be shown in detail below), and as we are
looking for the lower limit on $M_{\rm BH}$ we thus need to consider
Case~A.

\subsection{A progenitor model for Wray~977/GX~301-2}

The binary model (System No.~8; cf. Table~1) with initial parameters
selected in this way has a primary star initial mass of 26$\Msun$, a
secondary initial mass of 25$\Msun$, and an initial period of 3.5
days. The evolution of both components in the HR diagram is displayed
in Figure~\ref{fig:hrd}. It proceeds through Case~A mass transfer, for
which a rapid and a slow phase can be distinguished and which is
followed by a Case~AB mass transfer after the core hydrogen exhaustion
of the primary (cf. Fig.~\ref{fig:m_trans}).

\begin{figure*}[t]
\begin{centering}
\epsfxsize=0.8\hsize
\epsffile{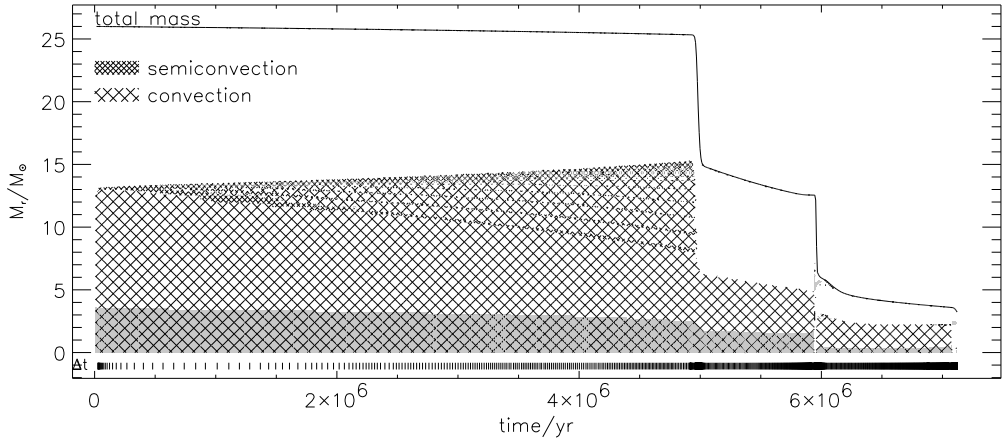}
\caption{Evolution of the internal structure of the primary
  component of System No.~8 --- i.e., our progenitor model for
  GX~301-2 --- as function of time.  Convection and semiconvection are
  marked as indicated, and gray shading designates regions of nuclear
  energy generation.  The upper solid line indicates the total mass of
  the star. The rapid phase of the Case~A mass transfer and the
  Case~AB mass transfer correspond to the sharp decreases of the total
  mass at about 4.9 and 5.9 Myr, respectively. Core hydrogen
  exhaustion coincides roughly with Case~AB mass transfer. Core
  helium burning ends at $t\simeq 7.1\,$Myr. The computations stops at
  core neon ignition. The decrease in mass between 4.9 and 5.9 Myr
  (about 2.5$\mso$) is due to slow Case~A mass transfer, that for
  $t\simgr 6\,$Myr (about 3.2$\mso$) due to WR winds.  The final mass
  of the star before it explodes as supernova is 3.17$\mso$
  (cf.~Table~1). }
\end{centering}
\label{fig:m26conv}
\end{figure*}

\begin{figure}
\epsfxsize=\hsize
\epsffile{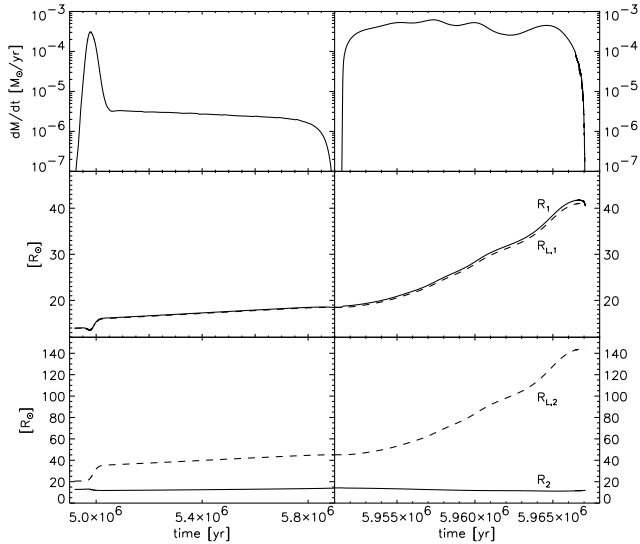}
\caption{Details of the mass transfer evolution of our progenitor
  model for Wray~977/GX~301-2, System No.~8. The left panel covers the
  Case~A mass transfer phase, the right panel the Case~AB.  Shown are:
  the mass transfer rate (upper panel), stellar radius $R_{\rm 1}$
  (solid line) and Roche radius $R_{\rm L,1}$(dashed line) of the
  primary (middle panel), and stellar radius $R_{\rm 2}$ (solid line)
  and Roche radius $R_{\rm L,2}$(dashed line) of the secondary (lower
  panel).}\label{fig:m_trans}
\end{figure}

The secondary evolves to a maximum mass of 42.4$\Msun$, which is
reduced by stellar winds later on. After the Case~AB mass transfer,
the primary is a helium star of 6.4$\mso$ (i.e., a Wolf-Rayet star)
which evolves to a final mass of 3.2$\Msun$ due to wind mass loss (cf.
Fig.~\ref{fig:m26conv}).  Note that the initial helium star mass of
this 26$\mso$ primary is smaller than the corresponding mass in a
25$\mso$ primary undergoing Case~B mass transfer (System No.~9 in
Table~1 below), since in Case~A primaries the hydrogen burning
convective core mass is reduced due to the mass transfer (cf.
Fig.~\ref{fig:m26conv}).

Our System No.~8 does not only represent the academic case which
yields the minimum initial progenitor mass of the neutron star
GX~301-2. It also fulfills all empirical constraints imposed by
Wray~977/GX~301-2 (cf. Section~1, and see Fig.~\ref{fig:hrd}) and is
thus a viable progenitor model for this system.  The mass of the
B~star at the time of the supernova explosion of the primary is
40.5$\mso$, and the final period of the system of 46.22~d before and
47.3~d after the supernova explosion --- without considering a
supernova induced kick on the neutron star --- is in good agreement
with the observed period. Note that the possibility of supernova kicks
render conclusions based on the observed period as difficult since the
observed eccentricity of $e=0.47$ is rather large; it only constrains
the period of the spherical orbit before the supernova explosion
approximately to the range 15$\,$d...59$\,$d.

\begin{figure}[t]
\epsfxsize=\hsize
\epsffile{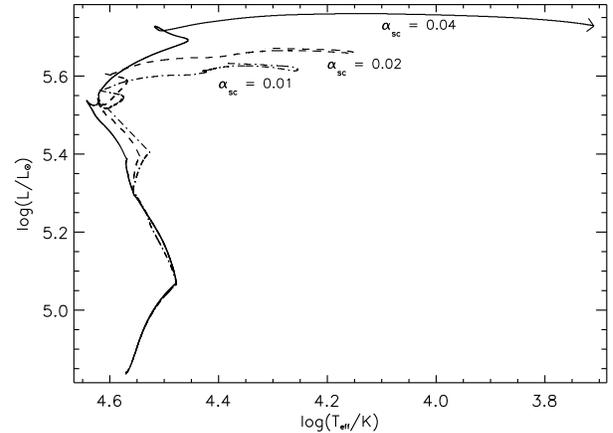}
\caption{Tracks in the HR diagram
  for the secondaries of the Systems No.~10
  ($\alpha_{\mathrm{sc}}=0.01$, dashed-dotted line), No.~10a
  ($\alpha_{\mathrm{sc}}=0.02$, dashed line), and No.~10b
  ($\alpha_{\mathrm{sc}}=0.04$, solid line), which have the same
  initial stellar masses of 25 + 24 $\Msun$ and the same initial
  period of 3.5~d (cf. Table~1). For Systems No.~10 and~10a, the
  tracks cover the evolution of the star from the zero age main
  sequence until the supernova explosion, reverse mass transfer does
  not occur. In System No.~10b, the secondary star rejuvenates during
  core hydrogen burning; reverse mass transfer would occur but is not
  taken into account in this calculation; the track ends before core
  helium ignition.  }\label{fig:m25hrd}
\end{figure}

While the luminosity of the mass gainer in our model~No.8 ($\llso
\simeq 5.7$) is within the observational error bar, it is slightly
more luminous than the value preferred by Kaper \&\ Najarro (1999)
($\llso \simeq 5.5$). To analyze the uncertainties of the post-main
sequence luminosity in our models, we have computed several $25 +
24\mso$ Case~A systems with different assumptions for the
semiconvective efficiency parameter $\alpha_{\mathrm sc}$.  This
parameter, which controls the so called rejuvenation process in the
accreting main sequence star (cf. Hellings 1983, Braun \& Langer
1995), has no influence on the evolution of the mass transfer, the
stellar masses or the binary period. However, it does affect the
evolutionary track of the secondary after the mass transfer. Most
important, it determines the temperature and radius evolution after
core hydrogen exhaustion, and to a smaller extent it influences the
stellar luminosity during this phase. From Figure~\ref{fig:m25hrd} we
see that smaller values of $\alpha_{\mathrm sc}$ lead to smaller
post-main sequence luminosities.  While the semiconvection parameter
has no relevance for our discussion of the critical mass limit for
neutron star formation, it is important for the probability to find
systems like Wray~977/GX~301-2 --- which is much higher for lower
$\alpha_{\mathrm sc}$ --- and for its future evolution (cf.
Fig.~\ref{fig:m25hrd}).  Note that System No.~8 has been computed with
an efficiency parameter for semiconvection of $\alpha_{\mathrm sc} =
0.02$ (see Table~\ref{fig:systems}).

We emphasize that our progenitor model for Wray~977 /GX~301-2, in
particular the initial mass of the neutron star progenitor as well as
its final CO-core mass, depend only very weakly on the convection
criterion or convective core overshooting.  This is particularly true
for the implication that the primary component forms a neutron star
since, in contrast to single stars, the final masses of the mass loser
in massive binaries (${\mathrm M > 25 \Msun}$) are determined by the
WR wind mass loss (see Sections~4.1 and~4.2, and
Figure~\ref{fig:mco-mi} below).

Finally, we want to mention that the surface composition of Wray~977
might be an additional way to discriminate evolutionary scenarios for
its progenitor evolution.  While according to scenarios which propose
a completely non-conservative evolution ($\beta =0$) the surface
chemical composition would be unaltered, our model predicts a
enrichment factor for nitrogen of 5.5 and depletion factors of carbon
and oxygen of 0.33, and 0.75, respectively. All isotopes of Li, Be and
B are totally depleted.

\subsection{Potential problems of the progenitor model}

Our progenitor model for Wray~977/GX~301-2 has two potential problems
which we want to discuss here for completeness, even though they may
turn out not to be essential.

The first concerns effects which the stellar wind of the secondary
star might have on the accretion efficiency parameter $\beta$.  In
principle, one could imagine a situation where the secondary's wind
drags part of the material which has left the primary and is on its
way to the secondary with it to infinity. However, detailed models for
such a situation are not available. We are optimistic that this effect
is not too important in our case since most of the mass transfer from
the primary to the secondary does not occur through an accretion disk,
but instead, according to the estimates of Ulrich \& Burger (1976),
the gas stream impacts directly onto the surface of the secondary.
This is so for the complete Case~A mass transfer as well as for the
ensuing Case~AB mass transfer in our System No.~8. Thus, the Case~A
binaries are those among the massive close binaries for which a
conservative evolution appears most appropriate.

The second potential problem is that for certain initial and physical
parameters the supernova order in Case~A systems can reverse, i.e.,
the secondary can explode before the primary does. We have shown in
Section~3.2 that Case~A models exist for which the supernova order is
not inverted (as needed to explain Wray~977/GX~301-2; e.g.  our model
No.~8). However, with $\alpha_{\rm sc}=0.01$ a reverse supernova order
would most likely occur in this system, as we conclude from our model
No.~10 (cf. Table~1). This does not mean that for $\alpha_{\rm
  sc}=0.01$ a supernova order reversal would occur for all
conservative Case~A systems; rather a slight decrease of the initial
secondary mass or a slight increase of the initial period would avoid
this to happen. It is therefore conceivable that for any value of
$\alpha_{\rm sc}$ the initial period-initial mass ratio parameter
space for Case~A systems for which the primary explodes first as a
supernova is not empty (cf.  also Pols 1994; Wellstein et al. 1999).

\subsection{Alternative progenitor scenarios for Wray~977}

During our literature search, we did not find any other evolutionary
calculations which were tailored to fit the system Wray~977/GX~301-2.

Several ``scenarios'' for the progenitor evolution of this system have
been considered. Brown et al. (1996), relying on the mass
determination of $\simgr 48\mso$ for Wray~977 by Kaper et al.  (1995),
prefer 45$\mso$ as the progenitor mass of GX~301-2.  They do not
consider constraints imposed by the binary period.  In Vanbeveren et
al.'s (1998a) scenario the system consists initially of a $40 +
36\mso$ pair in a 50~d orbit, which, in a non-conservative ($\beta =
0.5$) Case~B mass transfer, evolves into an $18.5 + 41\mso$ helium
star-O~star pair in a 28~d orbit. Interestingly, this is almost the
same configuration which results from the detailed evolutionary
calculations of de Loore \& De Greve (1992) starting also with a $40 +
36\mso$ pair but with a much smaller orbital period of 19.6~d.
Finally, Ergma \& van den Heuvel (1998) prefer a completely
non-conservative scenario ($\beta =0$) which implies an initial mass
of GX~301-2 in excess of $50\mso$, from which they draw far reaching
conclusions, in particular regarding the initial mass limit for black
hole formation.

In all of these scenarios, mass needs to be removed from the system.
However, the processes which would do this are not well understood.
Consequently, the amount of mass which leaves the system, and the
amount of angular momentum which is removed together with the mass are
not well constrained and need to be parameterized.  This shows, e.g.,
when the above mentioned results of Vanbeveren et al. (1998a) and de
Loore \& De Greve (1992) are compared, which differ in their
treatments of angular momentum loss.  Consequently, the period of
Wray~977/GX~301-2, or likewise the orbital separation of the two
components, can not be precisely predicted in these scenarios.  Only
the scenario of Ergma \& van den Heuvel (1998) does not include
binary-induced mass loss from the system and is thus free of this
uncertainty.

All of the above scenarios favor an initial mass of the neutron star
GX~301-2 in the range of 40...50$\mso$. Several of the papers
mentioned above conclude that therefore single stars with initial
masses in this range should form neutron stars as well. According to
our detailed Case~A evolutionary model presented in Section~3.2 such
conclusions can not be supported.  On the contrary, we find a likely,
and at least a possible, initial mass of GX~301-2 of only 26$\mso$.
Furthermore, as will be outlined in Section~5 below, this implies only
that single stars of initially 21$\mso$, not of 26$\mso$, form a
neutron star.  While this appears to be in comfortable agreement with
(admittedly uncertain) expectations from single star calculations
(e.g., Woosley \& Weaver 1995), neutron stars from 40...50$\mso$ stars
are not, which made Ergma \& van den Heuvel (1998) conclude that the
type of remnant of the mass loser --- neutron star or black hole ---
must depend on additional stellar parameters, for example magnetic
fields or rotation. Although we can not rule out that such effects
play a role, we can not conclude from our results that such effects
should be present.

\begin{table*}[tp!]
\caption{Key properties of the computed binary systems. Given are
the initial primary and secondary masses $M_{\rm 1,i}$ and $M_{\rm 2,i}$,
the initial mass ratio $q_{\rm i} = M_{\rm 2,i} / M_{\rm 1,i}$,
initial and final period $P_{\rm i}$ and $P_{\rm f}$, the maximum mass
of the secondary $M_{\rm 2,max}$, the mass transfer case, 
initial and final helium core mass of the primary $M_{\rm He,i}$
and $M_{\rm He,f}$ (note that the latter is equal to the final 
mass of the primary component for all considered cases), final
CO-core mass of the primary $M_{\rm CO,f}$, the central carbon
mass fraction of the primary at core helium exhaustion C$_{\rm c}$, the  
primaries final helium and carbon surface mass fraction
He$_{\rm s}$ and C$_{\rm s}$, and the amount of helium left in the
pre-supernova structure of the primary $\Delta M_{\rm He}$.
Models 2$^{\prime}$, 5$^{\prime}$,
7$^{\prime}$, 10$^{\prime}$, 15$^{\prime}$ and 17$^{\prime}$ are
computed with the WR-wind mass loss rates multiplied by a factor 
0.5 (Eq.~2 in Section~2), that for System No.~1$^{\prime\prime}$ with 0.25. 
No final periods have been derived for these systems.
All systems have been computed with a semiconvection
parameter of $\alpha_{\rm sc} = 0.01$ unless indicated otherwise.
System No.~8 corresponds to our progenitor model for the massive
X-ray binary Wray~977/GX~301-2 and is discussed in detail in Section~3.2}
\label{fig:systems}
\begin{tabular}{l c c c c l l c c l l c c c l}
\hline 
 ~ & $M_{\rm 1,i}$ & $M_{\rm 2,i}$ & $q_{\rm i}$& $P_{\rm i}$ & $P_{\rm f}$ &
$M_{\rm 2,max}$ & Case & $M_{\rm He,i}$ & $M_{\rm He,f}$ & $M_{\rm CO,f}$ &  
C$_{\rm c}$ & He$_{\rm s}$ & C$_{\rm s}$ & $\Delta M_{\rm He}$ \\ 
 & $\Msun$ & $\Msun$ & & d & d & $\Msun$ & & $\Msun$ & $\Msun$ & $\Msun$ & & & & $\Msun$ \\
\hline
1$^a$         & 60 & 34 &0.57& 20  & 59$^a$ & 58.9$^a$& B       & 26.8 & 3.13 & 2.35 & 0.35 & 0.41 & 0.48 & 0.24 \\
1$^{\prime\prime}$$^a$&60&34&0.57&20& ---    & --- &  B      & 26.8    & 7.55     & 5.93 & 0.25 & 0.14 & 0.47 & 0.12 \\
2             & 60 & 34 &0.57& 6.2 & 17.9   & 59.6 &A+AB     & 25.8 & 3.10     & 2.34 & 0.35 & 0.50 & 0.42 & 0.29 \\ 
2$^{\prime}$  & 60 & 34 &0.57& 6.2 & ---    & 59.6 &A+AB     & 25.8 & 4.07     & 3.07 & 0.32 & 0.28 & 0.53 & 0.17 \\
3             & 60 & 40 &0.67& 7.0 & 18.9   & 64.7 & A+AB    & 26.1 & 3.11     & 2.35 & 0.35 & 0.50 & 0.42 & 0.29 \\ 
4             & 46 & 34 &0.74& 5.0 & 18.0   & 55.9 & A+AB    & 18.2 & 3.10     & 2.34 & 0.35 & 0.48 & 0.44 & 0.27 \\ 
5             & 40 & 30 &0.75& 4.0 & 16.5   & 51.2 & A+AB    & 14.2 & 3.11     & 2.33 & 0.36 & 0.55 & 0.38 & 0.33 \\ 
5$^{\prime}$  & 40 & 30 &0.75& 4.0 & ---    & 51.2 & A+AB    & 14.2 & 3.84     & 2.87 & 0.33 & 0.34 & 0.51 & 0.20 \\
6$^a$         & 40 & 25 &0.63& 40  &130$^a$ &45.0$^a$& B     & 16.9 & 3.11     & 2.34 & 0.35 & 0.50 & 0.42 & 0.29 \\
7             & 30 & 24 &0.80& 4.0 & 22.0   & 41.9 & A+AB    & 9.92 & 3.12     & 2.33 & 0.36 & 0.59 & 0.36 & 0.36 \\ 
7$^{\prime}$  & 30 & 24 &0.80& 4.0 & ---    & 41.9 & A+AB    & 9.92 & 3.63     & 2.71 & 0.34 & 0.93 & 0.05 & 0.33 \\ 
{\bf 8}$^b$ &{\bf 26}&{\bf 25}&{\bf 0.96}&{\bf 3.5}&{\bf 46.2}&{\bf 42.4}&{\bf A+AB}&{\bf 6.41}&{\bf 3.17}&{\bf 2.33}&{\bf 0.37 }&{\bf0.98}&{\bf0.00}&{\bf 0.62} \\
9             & 25 & 24 &0.96& 5.0 & 34.9   & 39.2 & B       & 8.32 & 3.12     & 2.32 & 0.36 & 0.96 & 0.02 & 0.39 \\ 
10$^c$        & 25 & 24 &0.96& 3.5 & 44.1   & 40.6 & A+AB    & 6.02 & 3.19     & 2.36 & 0.37 & 0.98 & 0.00 & 0.66 \\
10a$^b$       & 25 & 24 &0.96& 3.5 & 47.6   & 40.6 & A+AB    & 6.06 & 3.19     & 2.31 & 0.37 & 0.98 & 0.00 & 0.71 \\
10b$^b$       & 25 & 24 &0.96& 3.5 &  ---   & 40.6 & A+AB    & 6.06 &   ---    &  --- &--- &  --- & ---  &  --- \\
10$^{\prime}$ $^c$&25&24&0.96& 3.5 &  ---   & 40.6 & A+AB    & 6.06 & 3.39     & 2.42 & 0.36 & 0.98 & 0.00 & 0.79 \\
11            & 25 & 19 &0.76& 4.0 & 35.7   & 35.9 & A+AB    & 6.27 & 3.18     & 2.32 & 0.37 & 0.98 & 0.00 & 0.71 \\ 
12            & 25 & 16 &0.64& 4.0 & 28.3   & 32.3 & A+AB    & 6.19 & 3.16     & 2.24 & 0.37 & 0.98 & 0.00 & 0.78 \\ 
13            & 22 & 18 &0.82& 3.0 & 49.2   & 34.2 & A+AB    & 4.48 & 3.13     & 1.90 & 0.37 & 0.98 & 0.00 & 1.12 \\ 
14            & 21 & 19 &0.90& 5.0 & 43.4   & 33.2 & B       & 6.00 & 3.14     & 2.03 & 0.36 & 0.98 & 0.00 & 0.98 \\ 
15            & 21 & 19 &0.90& 3.0 & 60.2   & 34.4 & A+AB    & 4.20 & 3.15     & 1.80 & 0.37 & 0.98 & 0.00 & 1.20 \\ 
15$^{\prime}$ & 21 & 19 &0.90& 3.0 &  ---   & 34.4 & A+AB    & 4.20 & 3.45     & 1.80 & 0.36 & 0.98 & 0.00 & 1.54 \\
16$^b$        & 21 & 19 &0.90& 3.0 & 57.7   & 34.5 & A+AB    & 4.23 & 3.07     & 1.73 & 0.35 & 0.98 & 0.00 & 1.25 \\
17            & 20 & 18 &0.90& 5.0 & 45.8   & 31.7 & B       & 5.56 & 3.15     & 2.17 & 0.37 & 0.98 & 0.00 & 0.73 \\ 
17$^{\prime}$ & 20 & 18 &0.90& 5.0 &  ---   & 31.7 & B       & 5.56 & 3.39     & 2.18 & 0.36 & 0.98 & 0.00 & 0.95 \\
18            & 20 & 18 &0.90& 3.5 & 55.2   & 32.5 & A+AB    & 4.30 & 3.00     & 1.69 & 0.37 & 0.98 & 0.00 & 1.23 \\ 
19            & 20 & 16 &0.80& 3.5 & 47.8   & 30.7 & A+AB    & 4.24 & 3.03     & 1.75 & 0.37 & 0.98 & 0.00 & 1.18 \\ 
20            & 20 & 14 &0.70& 3.5 & 48.4   & 28.9 & A+AB+ABB& 4.20 & 2.71     & 1.58 & 0.37 & 0.98 & 0.00 & 1.06 \\ 
21            & 20 & 18 &0.90& 2.5 & 65.5$^d$& 33.2 & A+AB+ABB& 3.55 & 2.7$^d$  & 1.56$^d$& 0.36 & 0.98 & 0.00 & 1.07$^d$ \\ 
22            & 19 & 17 &0.90& 5.0 & 46.7   & 30.2 & B       & 5.18 & 3.15     & 2.12 & 0.36 & 0.98 & 0.00 & 0.90 \\ 
23            & 18 & 17 &0.94& 5.0 & 49.3   & 29.5 & B       & 4.90 & 3.17     & 1.93 & 0.36 & 0.98 & 0.00 & 1.13 \\ 
24            & 18 & 16 &0.89& 5.0 & 45.7   & 28.6 & B       & 4.89 & 3.15     & 1.77 & 0.36 & 0.98 & 0.00 & 1.26 \\ 
25            & 18 & 16 &0.89& 3.0 & 167    & 29.7 & A+AB+ABB& 3.27 & 2.02     & 1.48 & 0.38 & 0.98 & 0.00 & 0.51 \\ 
26            & 16 & 15 &0.94& 8.0 & 172.1  & 25.7 & B+BB    & 3.83 & 2.32     & 1.51 & 0.36 & 0.98 & 0.00 & 0.74 \\ 
27            & 16 & 11 &0.69& 3.0 & 264$^d$& 24.2 & A+AB+ABB& 2.63 & 1.45$^d$ & 1.21$^d$& 0.40& 0.98 & 0.00 & 0.15$^d$ \\
28            & 13 & 12 &0.92& 3.1 &175$^d$ & 22.0 & B+BB    & 2.80 & 1.42$^d$ & 1.31$^d$& 0.40& 0.98 & 0.00 & 0.11$^d$ \\
\hline
\end{tabular}

$^a$ system was treated conservatively although a
   common envelope phase is expected. 
   The final period and the maximum secondary mass may thus be 
   greatly overestimated \\
$^b$ sequence was computed with $\alpha_{\rm sc}=0.02$,
  except for System No.~10b, where $\alpha_{\rm sc}=0.04$ was used \\
$^c$ reverse supernova order (no influence on core masses)\\
$^d$ masses are only upper ($M_{\rm He,f}$, $\Delta M_{\rm He}$) 
   or lower ($M_{\rm CO,f}$) limits, and the final
   period is only a lower limit, since the Case~BB or~ABB mass trasfer 
   phase was not finished at the end of the calculations. \\ 
\end{table*}

\section{The final stellar and core masses of massive primaries }
       
In this Section, we present the results of calculations for the
evolution of massive close binaries for a wide range of primary star
initial masses and for various initial system parameters.  The
evolution of the primary components is followed up to central neon
ignition in most cases.  Our goal is to derive final properties of the
primaries, in particular their final masses and core masses.  Those
are used in the next Section to constrain the critical initial mass
limit for neutron star/black hole formation in {\em single stars}. As
a by-product we also obtain results for the evolution of the surface
chemical composition of the primaries, which is relevant for the
understanding of Type~Ib/c supernovae; these issues are discussed in
Section~6.

We compute binary models for primary star initial masses from 13 to
60$\Msun$, and for various initial orbital periods and mass ratios.
We also study the differences between Case~A and Case~B mass transfer
for the resulting core masses. Since there are indications that the
mass loss rates for Wolf-Rayet stars (i.e., helium stars) have been
overestimated in the past (Hamann \& Koesterke 1998), we computed
sequences where our standard helium star mass loss rate (cf.  Section
2) has been multiplied by 0.5, in one case even by 0.25. An overview
over the grid of computed systems can be obtained from
Table~\ref{fig:systems}.

Our calculations are restricted to Case~A and early Case~B systems.
The latter are defined as such where the mass transfer occurs early
enough during the post-main sequence expansion of the primary that
contact is avoided.  The theory for the evolution of late Case~B and
Case~C systems is not yet very well developed; often it is assumed
that such system go through a common envelope evolution and either
merge --- in this case they are not relevant for the conclusions of
our work --- or expell the hydrogen-rich envelope of the primary
during the spiral-in phase of the secondary (e.g., Podsiadlowski
1992). Since this process also removes considerable amounts of orbital
angular momentum rather small orbital separations and periods are
expected from this type of evolution.

Three types of post-main sequence evolutionary tracks can be
distinguished. The first (say, Type~1) leads the star to the red
supergiant branch at the begining of core helium burning but allows
the stellar radius to increase after core helium exhaustion to values
which exceed those achieved earlier. Tracks like this are in fact
common, but the post-helium burning radius excess is mostly not very
large, in particular for stars with initial masses above 20$\mso$
(e.g., Schaller et al. 1992).  The second type of track (Type~2)
reaches the Hayashi-line only at the end of core helium burning.
However, this kind of evolution is found only for small metallicity
(cf., Schaller et al. 1992, Langer \& Maeder 1994), and the observed
large number of red supergiants discards it as a common case.  There
is also an intermediate type of evolutionary track (Type~3), where the
star reaches the Hayashi line for the first time {\em during} central
helium burning (e.g., the 20$\mso$ track at Z=0.02 or the 60$\mso$
track at Z=0.001 of Schaller et al. 1992).

Unfortunately, it is not possible at the present time to correctly
predict stars of which mass and metallicity follow which type of
track. The reason is that the temperature and radius evolution of
stars more massive than $\sim 10\mso$ is an extremely sensitive
function of internal mixing efficiencies (cf. Stothers \& Chin 1992,
Langer \& Maeder 1994) or mass loss rates (cf. Schaller et al 1992 and
Meynet et al. 1994), which suffer from large uncertainties. Note that
this problem must lead to large uncertainties in any prediction of the
number of black hole binaries, which is not always adequately
emphasized in population synthesis studies.  Since in the present
paper we deal only with stars of roughly solar metallicity, we neglect
the possibility of Type~2 tracks in the following discussion. We also
negect the unusual Type~3 tracks; however, as we can not prove that
they are not common, we note here that the initial-final mass relation
for primaries of this type would be inbetween that found for the
Case~A/B systems and that of single stars (cf. Fig.~5 below).

With these assumptions, the evolution of the primaries for systems
which do not merge or experience reverse mass transfer depends only on
the time during their evolution when the hydrogen-rich envelope is
removed.  Consequently, the primaries of late Case~B systems evolve
like primaries of early Case~B systems of the same initial mass.
Furthermore, as in Case~C systems the mass transfer starts only after
core helium exhaustion, and stellar wind mass loss after the mass
transfer can be neglected due to the short remaining stellar life
time, the helium cores of the primaries in these systems evolve like
the helium cores of single stars of the corresponding initial mass.

\subsection{Initial-final mass relations for primaries}

\begin{figure*}[p]
\begin{centering}
\vbox{
\epsfxsize=0.95\hsize
\epsffile{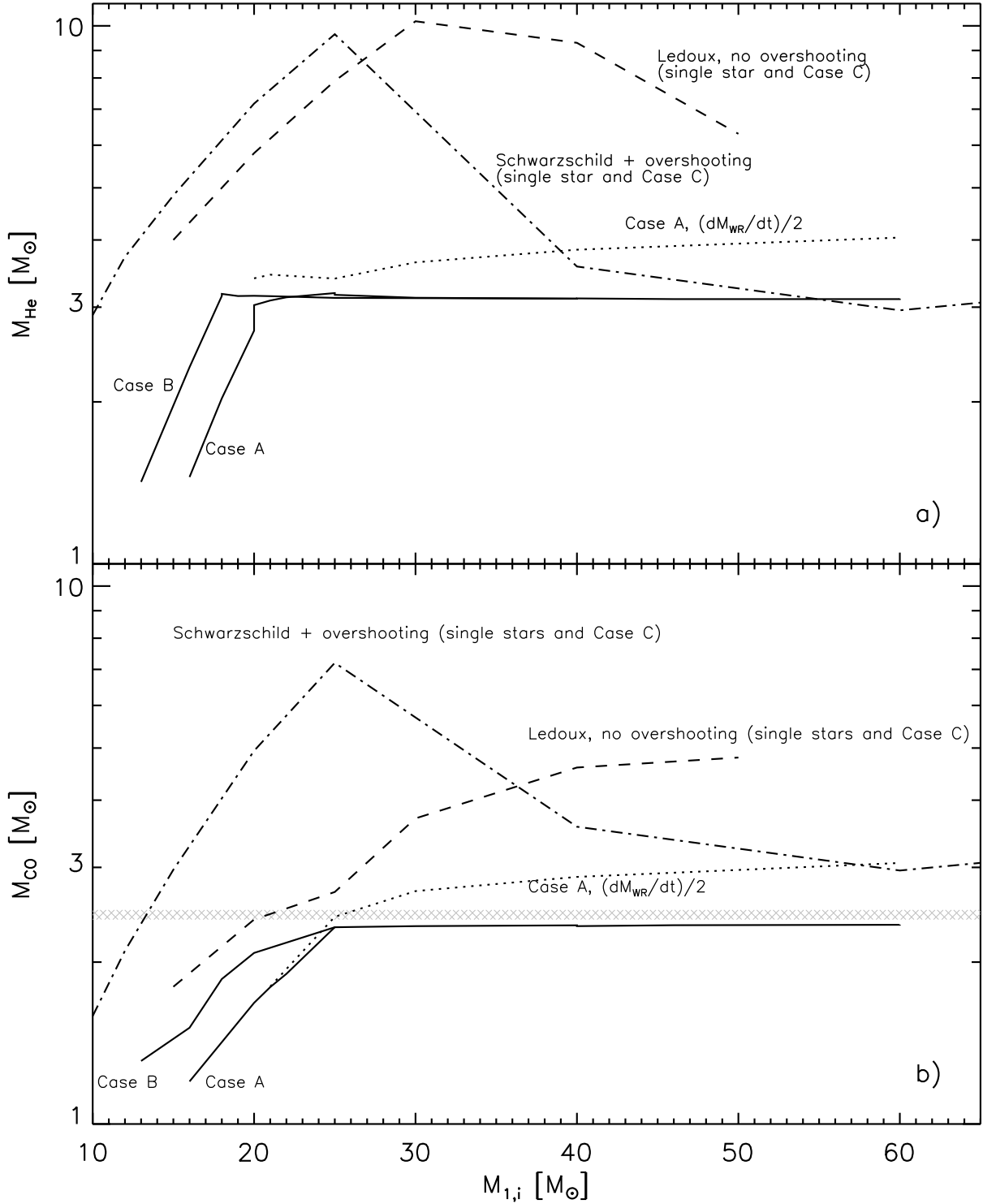}
}
\end{centering}
\caption{{\bf a)}
  Final He-core masses of the primaries of massive close binaries as
  function of their initial mass, for Case~A and Case~B systems (cf.
  Table \ref{fig:systems}) for nominal Wolf-Rayet mass loss (solid
  lines, below 25 $\Msun$ the upper line marks Case~B systems) and ---
  for the Case~A --- also for half the nominal Wolf-Rayet mass loss
  rate (dotted line), compared to the final helium core masses of
  single stars (dashed line; cf. Langer \& Henkel 1995; dotted-dashed
  line; cf. Maeder 1992) which may give a good approximation to the
  non-merging fraction of the Case~C systems.  {\bf b)}: The same as
  Fig.~\ref{fig:mco-mi}a, but for the final CO-core masses. The gray
  horizontal line marks the final CO-core mass of our progenitor model
  for the neutron star companion to Wray~977, GX~301-2}
\label{fig:mco-mi}
\end{figure*}

The results obtained for the initial-final mass relations of the
primaries in massive close binaries are shown in
Figure~\ref{fig:mco-mi}a. The data for Case~A and Case~B systems are
taken from Table~1, where the primaries final helium core mass is
designated as $M_{\rm He,i}$.  Since all of them are devoid of
hydrogen in their pre-supernova stage, the final helium core mass is
equal to their final mass. Except for Systems No.~21,~27 and~28, mass
loss occuring beyond the end of our calculations (neon ignition in
most cases) can safely be neglected due to the short remaining life
time of the star.

As expected from earlier work, the initial and therefore also the
final helium core masses of Case~B primaries do basically not depend
on the initial mass ratio or the initial orbital period (e.g., de
Loore \& De Greve 1992). Simply, the mass transfer stops only when
almost all of the hydrogen-rich envelope is removed from the primary.
However, since the period evolution does depend on the two mentioned
initial parameters, this simple logic does not apply any more to
primaries which evolve to final masses below $2.5...3\mso$, since
helium stars with masses smaller than this tend to evolve into red
giants (cf., Habets 1986) which then may or may not lead to a so
called Case~BB mass transfer (Delgado \& Thomas 1981).  We do not
investigate this process in detail here since it turns out that it
does not affect the initial mass limit for neutron star/black hole
formation. It is certainly important, though, for investigations of
the initial mass limit for white dwarf/neutron star formation in
binary systems.

Figure~\ref{fig:mco-mi}a shows that the initial-final mass relation of
Case~A systems differs from that of Case~B systems for initial primary
masses below $\sim 20\mso$. The reason is that Case~A mass transfer
leads to smaller initial helium core masses compared to Case~B, which
has been demonstrated in detail in Section~3.2 at the example of our
progenitor model for Wray~977. Note that there is no strict
initial-final mass relation for Case~A systems at all since the
initial helium core mass depends on time during core hydrogen burning
when the mass transfer starts.  Smaller initial helium star masses are
thus obtained for smaller initial periods (compare, e.g., systems
No.~18 and ~21). A dependence on the initial mass ratio seems to be
very small or absent (cf. Systems No.~18, 19, and~20).
 
That the initial-final mass relations for Case~A and~B primaries are
almost identical above $\simgr 25\mso$ is due to the fact that, for
our standard mass loss rate (Equation~1 in Section~2), the final
masses become independent of the initial helium star mass (Langer
1989). Even when the Wolf-Rayet mass loss rate is reduced by a factor
of~2 (Equation~2 in Section~2), the initial-final mass relation
remains rather flat, although a slight positive slope is found in this
case (Fig.~\ref{fig:mco-mi}a). Due to this mass convergence, also the
efficiency of convective core overshooting on the initial-final mass
relation above $\simgr 25\mso$ can be assumed to be small.
 
Fig.~\ref{fig:mco-mi}a shows also the initial mass-final helium core
mass relations for single stars obtained from the literature which, as
mentioned above, might also aply to Case~C mass transfer systems. It
is surprising that the treatment of convection, in particular the so
called convective core overshooting, introduces a much larger
uncertainty to the initial-final mass relation of single stars than to
that of Case~A/B binaries. The models of Langer \& Henkel (1995),
which have been computed with the same treatment of convection as the
models presented here, predict smaller final helium core masses for
stars below $\sim 30\mso$ but larger ones in the mass range
30...60$\mso$. For much higher initial masses, both sets of models can
be assumed to converge, as the curve derived from the models of Maeder
(1992) and that of the Case~A/B binaries do, due to the effect of mass
convergence (all computations use very similar Wolf-Rayet mass loss
rates).

\subsection{Initial mass - final CO-core mass relations}

In order to draw conclusions for the initial mass limit for the
formation of neutron stars/black holes, the initial-final mass
relation is not sufficient, since --- due to the strong mass loss,
particularly in binary systems --- stars with the same final mass or
even with the same final helium core mass may evolve different final
CO-core masses. This can be seen in Figure~\ref{fig:mco-mi}b, where we
plot the CO-core masses as function of the initial mass for the same
models as in Fig.~\ref{fig:mco-mi}a. E.g., the Case~A and~B primaries
with initial masses between 20~and 25$\mso$ end up with the same final
helium core mass but with largely different CO-core masses.

Above initial primary masses of $\sim 25\mso$, the final CO-core
masses are determined by the Wolf-Rayet winds.  As in the case of the
final helium core masses, the slope of the initial mass-final CO-core
mass relation is zero for the larger Wolf-Rayet mass loss rate and
slightly positive for the smaller mass loss rate. Again, the mass
transfer type (A~or~B) or the treatment of convection can be assumed
to have very little influence.

\subsection{Constraints from massive X-ray binaries}

Our 26$\mso$ progenitor for GX~301-2 (cf. Section~3.2) has a final
CO-core mass of 2.33$\mso$. Note that the initial mass-final CO-core
mass for Case~A/B primaries is particularly certain around this mass.
While uncertainties in the treatment of convection become relevant
only below $\sim 25\mso$ (cf. Section~4.2), significant differences in
the final CO-core masses due to differences in the Wolf-Rayet mass
loss rate occur only for higher initial primary masses.

As 26$\mso$ is the lowest possible initial mass for GX~301-2 we can
conclude that from Fig.~\ref{fig:mco-mi} that Case~A and~B primaries
with initial masses equal or less than this do not form black holes,
but instead evolve into neutron stars or white dwarfs.

On the other hand, the flatness of the initial mass-final CO-core mass
relation for Case~A/B primaries makes it very hard to form black holes
for {\em any} initial primary mass, and impossible to form high mass
black holes (cf. Section~5.2).  In fact, for the larger Wolf-Rayet
mass loss rate any black hole formation would be excluded.  For the
reduced Wolf-Rayet mass loss rate it may be possible to form black
holes, but it would be low mass black holes according to our
definition in Section~5.2 (cf. Fig.~\ref{fig:mco-mi}a). We computed
one massive system with a 60$\mso$ primary aplying only 1/4th of the
Wolf-Rayet mass loss rate (model No.~1$^{\prime\prime}$; cf. Table 1).
The primary ended as a 7.5$\mso$ Wolf-Rayet star.  A 36$\mso$ helium
core --- which may correspond to a 85$\mso$ star --- computed with the
same mass loss rate ended with 7.9$\mso$.

We conclude that, with the present stellar wind mass loss rates, it
appears to be impossible to explain a system like Cygnus~X-1 with a
black hole mass of $\sim 10\mso$ (Gies \& Bolten 1982, 1986; Herrero
et al. 1995) through Case~A or Case~B mass transfer.  Also the other
two potential massive black hole binaries, LMC~X-3 with a black hole
mass of $5.5\pm 1\mso$ (Kuiper et al. 1988), and LMC~X-1 with a
probable black hole mass of 6$\mso$ (Hutchings et al. 1987) are not
likely to be Case~A or~B remnants.  Only if the Wolf-Rayet mass loss
rates turn out to be substantially smaller than one fourth of Eq.~1, a
Case~A/B evolution for the progenitor of Cygnus~X-1 will be possible,
as suggested by Vanbeveren et al. (1998bc).  Anyway, a Case~C
evolution appears possible for this system.  E.g., for Cygnus~X-1,
Gies \& Bolten (1986) find that the mass of the optical component
should be $\simgr 20\mso$, while Herrero et al. (1995),by performing a
detailed spectral synthesis, find a most likely mass of 17.8$\mso$.
Thus, the progenitor of the black hole could have had an initial mass
$\simgr 25\mso$, which is consistent with a large black hole mass
according to Fig.~5a.

Interestingly, Brown et al. (1999) have recently proposed Case~C mass
transfer (mass transfer after central helium exhaustion) as a
possibility to obtain massive ($\sim 7\mso$; cf. Baylin et al. 1998)
black holes in the six known low mass black hole binaries; in fact,
due to the extrapolated large population of Galactic low mass black
hole binaries (Romani 1998), Portegies~Zwart et al. (1997) found
alternative explanations extremely difficult.

According to Fig.~\ref{fig:mco-mi}a, these black holes should then
come from an intermediate mass range, i.e. roughly 25...50 $\Msun$,
since the most massive Case~C primaries, like single stars, are
supposed to evolve through a Luminous Blue Variable phase (Langer et
al. 1994) where they lose the major part of their hydrogen-rich
envelope before core helium ignition and quickly transform into
Wolf-Rayet stars. They therefore would most likely avoid mass transfer
altogether (Vanbeveren 1991, Langer 1995).  According to
Fig.~\ref{fig:mco-mi}a, the maximum black hole initial mass in Case~C
systems, assuming that the hydrogen-rich envelope is completely
expelled at the end of the spiral-in process, is of the order of
10$\mso$. Lower stellar wind mass loss rates could lead to even higher
initial black hole masses.

From Figure~\ref{fig:mco-mi} it seems also conceivable that the
average black hole mass in black hole binaries is rather large, i.e.,
well above the final masses of Case~A/B primaries, as suggested by
Baylin et al. (1998). A distinction of two different regimes of
remnant masses has already been elaborated by Brown et al.  (1996).
While it can not be excluded that the most massive Case~A/B primaries
form black holes of relatively low mass ($\sim 3\mso$), their number
is not expected to be very large and is possibly zero
(Figure~\ref{fig:mco-mi}b).  On the other hand, assuming that in a
failed supernova the whole helium core forms the initial black hole
(cf. Fryer 1999) implies that most of those which form in Case~C
systems will have initial masses in excess of $\sim 5\mso$
(Figure~5a).

\section{Transforming the mass limit from binaries to single stars}

\subsection{Assumptions}

For single stars which do not lose their hydrogen-rich envelope
completely it is well known that the evolution of the stellar core is
decoupled from the envelope evolution. This allows the study of the
inner properties of stars during their late evolutionary phases ---
and thus also the investigation of their fate --- by just computing
the evolution of helium cores of a given mass (Arnett 1977, Woosley \&
Weaver 1988, Thielemann et al. 1996).

For stars which do lose their hydrogen-rich envelope well before core
helium exhaustion, i.e., in particular the primaries of massive close
binary systems, this approximation is not possible any more because
the mass of the helium core decreases during core helium burning.
However, as the mass loss never reaches the next inner core, the
CO-core, we will make the assumption in the following that the final
CO-core mass determines the fate of a massive star. In particular we
assume that CO cores with a mass below a critical value form neutron
stars or white dwarfs, while more massive CO-cores form black holes.

We note that the central carbon abundance at core helium exhaustion
$C_{\rm c}$ is in principle a futher independant parameter to be
considered.  It determines whether core carbon burning occurs
radiatively or convective, which affects the core entropy and the
final iron core mass (cf. Woosley \& Weaver 1995, Brown et al. 1999).
As more massive stars perform core helium burning at higher
temperatures, the general trend is that $C_{\rm c}$ is smaller for
higher initial masses.  There appears to be a critical carbon mass
fraction of about 15\% below which carbon burning is radiative (Weaver
\& Woosley 1993, Woosley \& Weaver 1995, Heger et al. 1999), and the
initial mass-final iron core mass-relation jumps discontinously to a
larger value (Timmes et al. 1996 Brown et al. 1999).  The situation is
complicated by the fact that $C_{\rm c}$ is not only dependant on the
initial stellar mass.  For stars which lose the hydrogen-rich envelope
during their evolution and uncover their helium core --- i.e. in
particular the primaries of massive close binaries ---, a higher mass
loss during core helium burning leads to larger values of $C_{\rm c}$
(Woosley et al. 1995, cf. also Table~1).

Even though Woosley \& Weaver (1995) find stars more massive than
$\sim 19\mso$ fall short of the critical carbon value of $\sim 15\%$,
recent models of Heger et al. (1999) which include effects of
rotational mixing find higher values of $C_{\rm c}$ for single stars
up to 25$\mso$.  Langer (1991) has shown that, for given initial mass
and mass loss history, $C_{\rm c}$ depends sensitively on assumptions
about convection (cf. Table~1 in Langer 1991).  Furthermore, it
depends on the still poorly known $^{12}$C($\alpha ,\gamma$) nuclear
reaction rate (Hoffman et al. 1999). Consequently, the values of
$C_{\rm c}$ must be considered as uncertain.

However, the central carbon mass fractions of our primaries are all
well above the critical value of $\sim 15\%$ (cf. Table~1). We can
thus assume a smooth CO-core mass-final iron core mass-relation for
our models (neglecting the statistical fluctuations in such a relation
due to silicon shell burning episodes; cf. Woosley \& Weaver 1995).
Furthermore, as all our models have extended convective core carbon
burning phases. As outlined by Brown et al. (1999) this results in
rather low core entropies and and relatively small iron core masses.
All together, we have reasons to assume that $C_{\rm c}$ is not an
essential independent parameter affecting the results of the present
study.

In principle, also further parameters might be important, i.e, the
stellar angular momentum or magnetic field; those are not considered
here.

\subsection{Results}

The approximation often made for single stars that the fate depends
monotonously on the initial stellar mass is not generally applicable
for close binary components.  This can be seen in
Figure~\ref{fig:mco-mi}b, which shows that the final CO-core mass (and
thus the fate) for a given initial mass depends on the previous mass
transfer evolution.

The progenitor of the neutron star in Wray 977/GX~301-2 must have had
a CO-core mass of at least 2.33$\mso$, corresponding to the minimum
initial mass of the primary of 26$\mso$.  Fig.~\ref{fig:mco-mi}b shows
that single stars above 21$\mso$ form CO-cores of at least 2.33$\mso$.
This allows us to derive a minimum black hole progenitor mass for
single stars of 21$\mso$ from Fig.~\ref{fig:mco-mi}b. This initial
mass limit refers to models computed with the Ledoux criterion for
convection (Langer \& Henkel 1995).  For single stars computed with
the Schwarzschild criterion and overshooting (Maeder 1992) we derive a
limit of 13 $\Msun$ from the same arguments. Since the two assumptions
about convection can be considered as the two extreme cases, we
conclude that single stars with initial masses below 13...21$\mso$
form neutron stars.

We want to emphasize that this is the strongest statement that, at the
present time, can be derived from the existence of a neutron star
companion to Wray~977. In particular, even though especially the lower
limit of 13$\mso$ appears disappointingly un-constraining, we must
consider arguments in the literature which derive much higher initial
mass limits for neutron star formation from the system
Wray~977/GX~301-2 as wrong.

If the hitherto poorly investigated Case~C mass transfer in massive
stars would always lead to the merging of both components, the
presence of black holes in massive binaries, e.g. that in Cyg~X-1,
would exclude the standard WR wind mass loss rates (dotted lines in
Fig.~\ref{fig:mco-mi}), because using this we find no systems with
more massive final CO-cores than for the neutron star progenitor of
GX~301-2 (cf. also Ergma \&\ van den Heuvel 1998). For the reduced WR
wind mass loss rate --- assuming that at least some of the mass losers
in binaries reach massive enough CO-cores to collapse to a black hole
--- we derive an upper CO-core mass limit up to which neutron star
formation might be possible of 3$\mso$. This would correspond to a
single star initial mass limit of 26$\mso$ with the Ledoux criterion
and of 15$\mso$ with the Schwarzschild criterion and overshooting.
These are lower mass limits for black hole formation in single stars
{\em if}~Case~A/B binaries could be shown to produce any black hole at
all.

It is interesting to distinguish the delayed black hole formation due
to fall back (e.g., Woosley \& Weaver 1995), thermal effects (Wilson
et al. 1986, Woosley \& Weaver 1986), or effects of the equation of
state (Brown \& Bethe 1994), and the prompt formation of black holes.
While delayed black hole formation gives rise to a ``normal''
supernova explosion, prompt black hole formation does not, although
also in this case the hydrogen-rich envelope may be ejected (MacFadyen
\& Woosley 1999, Fryer 1999).  In the first case, the black hole mass
equals the mass of the initially formed neutron star plus the mass of
the matter which falls onto it later-on. Thus, black holes of rather
low mass may form like this (Brown \& Bethe 1994, Brown et al. 1996).
For the prompt stellar collapse to a black hole, it may be a
reasonable assumption that the black hole mass roughly equals the
final helium core mass (MacFadyen \& Woosley 1999, Fryer 1999). Even
though these assumptions are uncertain and their confirmation must
await a better understanding of the core collapse supernova mechanism,
it probably makes sense that promptly formed black holes have larger
masses than those formed through the delayed scenarios (Brown \& Bethe
1994, cf., however, Woosley \& Weaver 1995).  In the line of Brown \&
Bethe (1994), we will thus consider the first as high mass black holes
(HMBHs) with the mass of the progenitor's helium core mass, and the
latter as low mass black holes (LMBHs) with a mass of $\simle 3\mso$.

The observation that Supernova~1987A, which had a progenitor of $\sim
20\mso$, did not promptly form a black hole (Arnett et al. 1989) can
give important constraints on the black hole formation limits.
Considering the single star models computed with the Schwarzschild
criterion and overshooting, we conclude from Fig.~5b that prompt black
hole formation can only occur in CO-cores more massive than $\sim
5\mso$.  This limits the formation of high mass black holes in single
stars or Case~C binaries to the initial mass range 20...32$\mso$.
Masses of high mass black holes of the range 5...10$\mso$ could be
produced (Fig.~5a). Case~A/B binaries could not form high mass black
holes, and even the production of low mass black holes might be
excluded as the core mass range which is able to yield to delayed
black hole formation is rather narrow (cf. Brown et al. 1999). If the
existence of a neutron star in the remnant of SN~1987A were confirmed
(cf., Wu et al. 1998, Fryer et al. 1999) the models using the
Schwarzschild criterion and overshooting might have a problem, since
only a very small initial mass range might be left to form high mass
black holes.

For models using the Ledoux criterion for convection, these mass
limits work out quite differently. From the neutron star companion to
Wray~977, which resulted in a single star initial mass limit for black
hole formation of $> 21\mso$ (see above), it follows that SN~1987A
formed a neutron star, in accord with current models and observations
(Wu et al. 1998, Fryer et al. 1999).  No further firm initial mass
limit can be derived from Fig.~5b, neither for low nor for high mass
black hole formation.  However, as stars above 21$\mso$ form much more
massive CO-cores than stars around 20$\mso$, in particular stars with
initial masses above $\sim 30\mso$ (Fig.~5b), it is conceivable that
single stars and Case~C primaries above $\sim 30\mso$ or so form high
mass black holes.  As for the models computed with the Schwarzschild
criterion, Case~A/B binaries would not produce high mass black holes.

For the case that {\em all} black holes in close binaries form through
the Case~C channel, we can only derive a less stringent lower mass
limit for black hole formation as for the case that {\em some} black
holes from in Case~A/B binaries.  The limit then comes from the fact
that the Case~C, at solar metallicity, is restricted to initial
primary masses of less than $\sim 40\mso$ (cf. Sect.~4.3).  The large
number of low mass black hole binaries (Portegies Zwart et al. 1997,
Romani 1998) with massive black holes ($\sim 7\mso$; cf. Baylin et al.
1998) makes it plausible that single stars in the mass range 25...40
$\Msun$ form high mass black holes.

Thus, independent of whether black hole binaries form preferentially
through Case~A/B or Case~C, we conclude that very likely single stars
with initial masses above $\sim 25\mso$ form black holes. This value
is in agreement with the recent result from core collapse simulations
of Fryer (1999).

\section{The structure of supernova progenitors}

The massive close binary models discussed in the previous Sections do
not only give predictions for the final stellar masses of both
components, but they also yield the final mechanical and chemical
structure of their envelopes.  While those primaries of our systems
which do not collapse directly to black holes become Type~Ib or~Ic
supernovae, the final explosions of the secondaries are hydrogen-rich
and have thus to be classified as Type~II supernovae (cf. Langer \&
Woosley 1996).
 
The mechanical structure of the secondaries of massive close binaries
can differ appreciably form single star Type~II supernova progenitors.
While the latter are red supergiants, our secondaries are typically
blue supergiants in their final stages. Whether or not massive
secondaries explode as blue or red supergiants depends critically on
the semiconvective mixing efficiency during the accretion phase (cf.
Fig.~4), i.e., whether or not the secondary rejuvenates (Hellings
1983, Braun \& Langer 1995). In fact, for the the semiconvective
mixing efficiency adopted here we can estimate a number ratio of
Type~II supernovae from blue supergiants to Type~Ib/c supernovae of
order unity.  This does not contradict the fact that many more
Type~Ib/c supernovae are observed, since exploding blue supergiants
are about 4~magnitudes dimmer than Type~Ib supernovae (Young \& Branch
1989), as demonstrated by Supernova~1987A (Arnett et al. 1989).

\begin{figure}[t]
\epsfxsize=\hsize
\epsffile{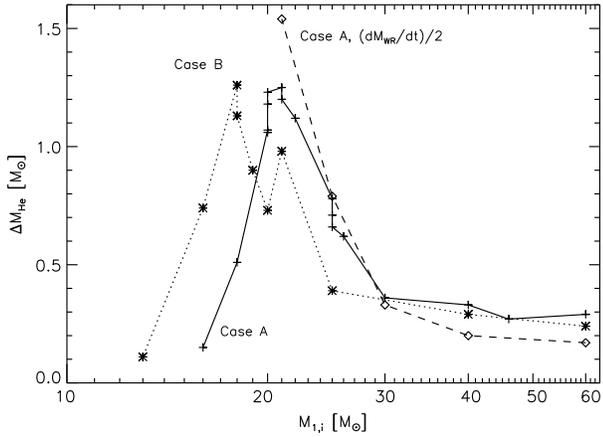}
\caption{The amount of helium present in the primary components
  at their immediate pre-supernova stage as function of their initial
  mass, for the computed Case~A (solid line) and Case~B (dotted line)
  systems.  The dashed line corresponds to Case~A systems computed
  with a WR wind mass loss rate reduced by 50~\%.}
\label{fig:dmhe}
\end{figure}

As mentioned above, the primaries of the systems studied in this work
will mostly produce Type~Ib/c supernovae. As a main criterion to
distinguish Type~Ib and Type~Ic supernovae is the detection of helium
lines in Ibs but their absence in Ics (Harkness \& Wheeler 1990), we
list in Table~\ref{fig:systems} the remaining mass of helium in the
envelope of the primaries at the time of their supernova explosion.
Figure~\ref{fig:dmhe} shows this quantity plotted as a function of the
primaries initial mass, separately for Case~A and Case~B systems. It
shows that substantial amounts of helium (i.e. more than $0.5\mso$)
remain only in a limited initial mass range, i.e. above $\sim 15\mso$
and below $\sim 25\mso$. Progenitors with lower or higher initial
masses, although they do not become helium free, may end up with as
little as $\sim 0.1\mso$ of helium.
  
For progenitors below $\sim 15\mso$, the reason for the low amounts of
helium in the pre-supernova structure is the occurrence of a Case~BB
or Case~ABB mass transfer as the low mass primaries, at the end of
their helium-star phase, extend to red giant dimensions (cf.
Section~4.1).  These objects end with a total mass very close to the
Chandrasekhar mass.  For the Case~A, this happens in our models for
primary initial masses below 18 $\Msun$, although even the 20 $\Msun$
primaries undergo a brief Case~ABB mass transfer which is, however,
too short to remove significant amounts of mass. A comparable
situation occurs in System No.~26, a Case~B system with a $16\mso$
primary: the Case~BB mass transfer occurs but stops before the major
part of the helium envelope is removed.  As it may have observational
implications, we note that some of these objects explode {\em during}
the mass transfer.

The Type~Ib/c progenitors with initial masses above $\sim 25\mso$ are
helium-poor due to strong stellar winds during a Wolf-Rayet phase.
Figure~\ref{fig:dmhe} shows that this result is not strongly affected
by uncertainties in the assumed Wolf-Rayet mass loss rate.  {\em
  Smaller} amounts of helium remain for the most massive progenitors
when the mass loss rate is reduced by a factor of~2.  The reason is
that the lower Wolf-Rayet mass loss leads to larger convective cores
in the advanced helium burning stages.  The final masses of these
stars is well above the Chandrasekhar mass.

Although the remaining amount of helium is similar for initial primary
masses below $\sim 15\mso$ and above $\sim 25\mso$, the envelope mass,
i.e. the amount of mass above the CO-core, is much larger in the
latter. While the stars from below $\sim 15\mso$ have a mantle of pure
helium (plus 2\% metals) on top of their CO-core, the envelope of the
primaries from above $\sim 25\mso$ is strongly enriched with carbon
and oxygen, as can be seen from the final surface abundances shown in
Table~1.

We could now speculate that the Type~Ic supernovae correspond to
Case~A/B primaries with initial masses below $\sim 15\mso$ and above
$\sim 25\mso$. As single stars above $\sim 40\mso$ evolve into
Wolf-Rayet stars with correspondingly low remaining amounts of helium
in the pre-supernova stage, those, together with Case~C primaries
above $\sim 40\mso$, might as well contribute to the Type~Ic supernova
class. In fact, small amounts of helium might still be compatible with
Type~Ic supernovae (Filippenko et al. 1995).  On the other hand,
Case~A/B primaries from the initial mass range 15$\mso$...25$\mso$ as
well as Case~C primaries from initial masses below $\sim 40\mso$ might
evolve into Type~Ib supernovae.

On the other hand, Woosley \& Eastman (1995) argued that the amount of
helium seen in a Type~Ib/c supernova (i.e., the helium line strengths)
may not only depend on the amount of helium which is present. They
conclude that it is essential whether radioactive $^{56}$Ni is mixed
close to or into the helium layer during the explosion, as the nickel
decay can excite the helium line. Hachisu et al. (1991) have shown
that more such mixing is expected in the explosions of lower mass
helium stars.

In summary, close binary primaries with the largest initial masses
($\simgr 25\mso$ for the Case~A/B) as well as single stars and wide
binaries (denoted ``Case~C'', although no mass transfer occurs; cf.
Section~4.3) with initial masses above $40...50\mso$ are the best
candidates for Type~Ic supernova progenitors. In their pre-supernova
stage, they contain low amounts of helium in their envelopes {\em and}
they might not experience significant mixing of radioactive $^{56}$Ni
into the helium layer.  This is interesting in context with the
peculiar Type~Ic supernova~1998bw and its associated weak $\gamma$-ray
burst (Kulkarni et al. 1998). In the collapsar model of MacFadyen \&
Woosley (1999), $\gamma$-ray bursts may be formed in helium cores
which evolve a sufficiently massive iron core to undergo a prompt
collapse to a black hole, given that it has the right amount of
angular momentum.  Independent of the convection physics, a prompt
black hole formation {\em and} a Type~Ic supernova with small amounts
of helium appear possible for Case~C binaries and also for single
stars with initial masses of $\sim 40\mso$ or above.  It is thus
consistent with our conclusions that SN~1998bw is a borderline case of
a successful $\gamma$-ray burst produced according to the model of
MacFadyen \& Woosley (1999).

\section{The overall picture}

In Table~2, we summarize a possible overall picture for the outcome of
massive close binary evolution through the various mass transfer
cases, and for single star evolution.  One can, however, {\em not}
pick initial primary mass and period (i.e., mass transfer case) and
then obtain the final answer at the corresponding position in the
table. Due to uncertainties in the convection efficiency, the
Wolf-Rayet mass loss rate, the post-main sequence radius evolution of
massive stars, and the CO-core-compact remnant mass relation, the
dividing lines between the various entries in Table~2 can not yet be
completely determined.

Table~2 should rather be regarded as a diagram with the initial binary
period as the x-axis and the initial primary or stellar mass as
y-axis. Several {\em qualitative} trends can be found in this
``diagram'', which should be taken more seriously than the exact
location of the dividing lines.

\begin{table}[t]
\caption{Type of remnant of primary components of massive close
binaries, i.e., white dwarf (WD), neutron star (NS), low 
(LMBH) or high mass black hole (HMBH), and expected supernova
type (if an explosion occurs, which is not secure for HMBHs)
as function of the primary initial mass and the mass transfer type
for a possible scenario of massive close binary evolution.
The initial mass ranges are not to be taken too literally
(see text). }
\label{fig:cc}
\begin{tabular}{ l  l l l l }
\hline
\noalign{\vskip 3pt}
M$_{1, \rm i}$  & Case A & Case B & Case C & single star \\
$\mso$          &        &      &        &               \\
\noalign{\vskip 3pt}
\hline
\noalign{\vskip 5pt}
8...13  &   WD     &    WD         &   SN~Ib    &    SN~II   \\
        &          &               &    NS      &     NS     \\
\noalign{\vskip 15pt}
13...16 &   WD     & SN~Ib or~Ic   &   SN~Ib    &    SN~II   \\
        &          &   NS          &     NS     &      NS    \\
\noalign{\vskip 15pt}
16...25 &  SN~Ib   & SN~Ib         &   SN~Ib    &    SN~II   \\
        &    NS    &     NS        &     NS     &     NS     \\
\noalign{\vskip 15pt}
25...40 &  SN~Ic   & SN~Ic         &   SN~Ib    &    SN~II   \\
        &   NS     &   NS          &    HMBH$^a$&    HMBH$^a$\\
\noalign{\vskip 15pt}
$> 40$  &  SN~Ic   & SN~Ic         &   SN~Ic    &    SN~Ic   \\
        & NS/LMBH  & NS/LMBH       &   HMBH$^b$ &   HMBH$^b$ \\
        &          &               &NS/LMBH$^c$ &NS/LMBH$^c$ \\  
\noalign{\vskip 3pt}
\hline
\end{tabular}

$^a$ mass range depends on assumptions on convection; see text\\
$^b$ for models computed with the Ledoux criterion for convection\\
$^c$ for models computed with the Schwarzschild criterion and 
     convective core overshooting\\
\end{table}

Most important, we see that the dividing lines between equal final
stages are not horizontal but sloped, with a positive slope (for
larger initial masses having smaller y-values) for all of them. For
example, the dividing line between white dwarf and neutron star
formation is (roughly) at 16...13$\mso$ for Case~A binaries (cf.
System No.~27 in Table~1) --- it depends smoothly on the initial
period for Case~A as shown by Wellstein et al. (1999) ---, at $\sim
13\mso$ for the Case~B (cf. System No.~28 in Table~1), and somewhere
between~8 and~10$\mso$ for single stars and Case~C binaries.
Similarly, high mass black holes are preferentially produced for large
initial masses {\em and} large initial periods.  Similar trends can be
found for the two dividing lines between Type~Ib and~Ic supernovae
(where the one at smaller initial masses may or may not correspond to
the Type~Ib/Ic distinction; cf.~Section~6).

This slope of the dividing lines between equal final stages are not
small. For example, our models are consistent with an initial primary
mass limit for black hole formation in Case~A/B systems of well above
$100\mso$, while in Case~C systems and single stars we derive a value
of $\sim 25\mso$.  We thus emphasize that this effect should be
included in population synthesis models which predict X-ray binary or
$\gamma$-ray burst frequencies. For the same reason, the neutron star
or black hole mass functions derived from single star models can not
simply be compared with mass functions of compact companions in X-ray
binaries.

As discussed in Section~5.2, the predicted initial mass range for high
mass black hole formation in Case~C binaries and single stars is quite
different for models using the Ledoux criterion for convection
($\simgr 30\mso$) compared to models using the Schwarzschild criterion
and overshooting (20...32$\mso$). In the latter case, the most massive
single stars would evolve into low mass black holes or even neutron
stars.  This conclusion can be changed only if the Wolf-Rayet mass
loss rate were smaller than one fourth of that in Eq.~1 (cf.
Vanbeveren et al. 1998b).

Finally, we note that the orbital periods of the Case~A,B,C binaries
at the time when the primary has terminated its evolution occur in
reverse order as the initial orbital periods (cf., Verbunt 1993).
Early Case~B systems end with shorter periods than the Case~A systems
because, for a given primary initial mass, less mass is transfered
during a Case~B mass transfer compared to the amount of mass which is
transfered in a Case~A and~AB together.  Furthermore, while the
conservative Case~A and~B binaries evolve to periods of several weeks
and months --- our Wray~977/GX~301-2 progenitor model No.~8 with a
final period of 46~days is a typical case --- late Case~B and Case~C
systems lose substantial amounts of mass and angular momentum and thus
become short period binaries (if they do not merge). E.g., the system
4U~1700-37, a $\sim 30\mso$ O~star with a 2.6$^{+2.3}_{-1.4}\mso$
compact companion and a 3.4~day orbital period (Heap \& Corcoran 1992,
Rubin et al. 1996), must have gone through a non-conservative mass
transfer.  This implies that the progenitor mass of the compact object
is likely to be larger than that of GX~301-2.

Consequently, conservative Case~A and~B systems, which are modeled in
detail in the present work, are the best candidates for progenitors of
close binaries consisting either of a helium star and a compact
remnant or of two compact stellar remnants --- i.e. for binary
$\gamma$-ray burst progenitors (cf. Fryer et al. 1999, and references
therein): Due to the large periods at the end of the primary's
evolution the binding energy of the secondary's envelope is low when
the reverse mass transfer occurs, which enhances the chance for the
ejection of the envelope during the spiral-in of the compact remnant
of the primary. In short period binaries like 4U~1700-37, a merging of
both components is rather more likely (cf. Podsiadlowski et al. 1995).

The known black holes X-ray binaries which possibly all have high mass
black hole companions (cf. Ergma \&\ van den Heuvel 1998, Baylin et
al. 1998, Brown et al. 1999) have such low periods that none of them
could have undergone a conservative evolution.  The fact that
non-conservative evolution occurs for all Case~C systems but not for
all Case~A or~B systems is in agreement with the origin of high mass
black holes in binaries as due to Case~C mass transfer (cf. Table~2),
as proposed by Brown et al. (1999). The predominance of short periods
in massive black hole binaries may be due to selection effects.  On
the other hand, neutron star binaries with much longer periods,
consistent with conservative evolution, have been found (e.g.
Wray~977/GX~301-2)). This is again consistent with the evolutionary
scenario depicted above.

It is important to note that Table~2 has been derived only for stars
of about solar metallicity. The radius evolution of massive stars, and
possibly also the Wolf-Rayet mass loss rates, depend strongly on
metallicity. Also, we do not work out the relative frequencies with
which the evolution through the various mass transfer channels occur.
E.g., at solar metallicity, the Case~C evolution may be relatively
rare (cf. Section~4), which could be the reason why only one Galactic
high mass X-ray binary is know to contain a high mass black hole
(Cyg~X-1), compared to two in the Large Magellanic Cloud. Furthermore,
Wellstein et al. (1999) find that conservative evolution for Case~B
systems is restricted to initial primary masses below $\sim 25\mso$,
while Case~A systems are more likely to avoid contact for initial
primary masses above $\sim 15\mso$. All this may change for different
metallicities. To work this out has to be left to future
investigations.

\begin{acknowledgements}
  We are very grateful to Hans Bethe, Gerry Brown, Chris Fryer,
  Alexander Heger, Lex Kaper, Chang-Hwan Lee, and Stan Woosley for
  stimulating discussions and for the communication of results prior
  to publication.  This work has been supported by the Deutsche
  Forschungsgemeinschaft through grants La~587/15-1 and 16-1.
\end{acknowledgements}

\end{document}